\documentclass[11pt]{article}
\usepackage{color}
\usepackage{amsmath,amssymb}
\usepackage{graphicx} 
\usepackage{bbm}
\usepackage{longtable}
\usepackage[small,loose]{subfigure}
\usepackage[small]{caption2} 
\usepackage{fleqn} 
\usepackage{cite}
\usepackage{pifont}

\advance \headheight by 3.0truept       
\setcaptionwidth{0.75\textwidth}

\newcommand{\E}[1]{\ensuremath{\mathrm{E}_{#1}}} 

\newcommand{\SO}[1]{\ensuremath{\mathrm{SO}(#1)}}
\newcommand{\SU}[1]{\ensuremath{\mathrm{SU}(#1)}}
\newcommand{\U}[1]{\ensuremath{\mathrm{U}(#1)}}
\newcommand{\Z}[1]{\ensuremath{\mathbbm{Z}_{#1}}} 

\allowdisplaybreaks

\begin{document}

\date{}
\title{\begin{flushright}
\ \\*[-80pt]
\begin{minipage}{0.3\linewidth}
\normalsize
CERN-PH-TH/2006-230\\
OHSTPY-HEP-T-06-006 \\
NSF-KITP-06-100 \\
TUM-HEP-651/06 \\*[50pt]
\end{minipage}
\end{flushright}
{\bf\huge A Mini--Landscape of Exact MSSM Spectra in
Heterotic Orbifolds}\\[0.8cm]}

\author{{\bf\normalsize
Oleg~Lebedev$^{1,2}$\!,
Hans~Peter~Nilles$^1$\!,
Stuart~Raby$^3$\!,
Sa\'ul~Ramos-S\'anchez$^1$\!,}\\{\bf\normalsize
Michael~Ratz$^4$\!,
Patrick~K.~S.~Vaudrevange$^1$\!,
Ak{\i}n~Wingerter$^3$}\\[1cm]
{\it\normalsize
${}^1$ Physikalisches Institut der Universit\"at Bonn,}\\[-0.05cm]
{\it\normalsize Nussallee 12, 53115 Bonn,
Germany}\\[0.2cm]
{\it\normalsize
${}^2$ CERN, Theory Division, CH-1211 Geneva 23, Switzerland}\\[0.2cm]
{\it\normalsize
${}^3$ Department of Physics, The Ohio State University,}\\[-0.05cm]
{\it\normalsize
191 W.\ Woodruff Ave., Columbus, OH 43210, USA}\\[0.2cm]
{\it\normalsize
${}^4$ Physik Department T30, Technische Universit\"at M\"unchen,}\\[-0.05cm]
{\it\normalsize James-Franck-Strasse, 85748 Garching,
Germany}
}
\maketitle \thispagestyle{empty}

\begin{abstract}
{ We explore a ``fertile patch'' of the heterotic landscape based on
a \Z6-II orbifold with SO(10) and \E6 local GUT structures. We
search for models allowing for the exact MSSM spectrum. Our result
is  that of order 100  out of a total $3 \times 10^4$ inequivalent
models  satisfy this requirement. }
\end{abstract}
\clearpage

\section{Introduction}

Although there are only a few consistent 10D string theories, there is a  huge
number of 4D string compactifications \cite{Lerche:1986cx,Bousso:2000xa}.  This
leads to the picture that string theory has a vast landscape of vacua
\cite{Susskind:2003kw}. The (supersymmetric) standard model (SM) corresponds to
one or more possible vacua which  \emph{a priori}
might not be better than others. To
obtain  predictions from string theory one can employ the following
strategy: first seek vacua that are consistent with observations and then study
their properties. Optimistically, one might hope to identify certain features
common to all realistic vacua, which would lead to predictions. Even if this is
not the case, one might still be able to assign probabilities to certain
features, allowing one to exclude certain patches of the landscape on a statistical
basis.  However, realistic vacua are very rare. For instance, in the context of
orientifolds of Gepner models, the fraction of models with the chiral matter
content of the standard model  is about $10^{-14}$
\cite{Dijkstra:2004cc,Anastasopoulos:2006da}. The probability of getting
something close to the MSSM in the context of intersecting D--branes in an
orientifold background is $10^{-9}$ \cite{Gmeiner:2005vz,Douglas:2006xy}, even
if one allows for chiral exotics. In this study, we show that certain patches of
the heterotic landscape are more ``fertile'' in the sense that  the analogous
probabilities are at the percent level.

We base our model scan on the heterotic $\E8\times\E8$ string
\cite{Gross:1984dd,Gross:1985fr}
compactified on an orbifold
\cite{Dixon:1985jw,Dixon:1986jc,Ibanez:1986tp,Ibanez:1987xa,Ibanez:1987sn,
 Casas:1988hb,Casas:1987us}. 
Our study is motivated by recent work on an orbifold GUT
interpretation of heterotic string models
\cite{Kobayashi:2004ud,Forste:2004ie,Kobayashi:2004ya}.
We focus on the \Z6-II orbifold, which is described in detail in
\cite{Kobayashi:2004ud,Kobayashi:2004ya,Buchmuller:2006ik}. The search strategy
is based on the concept of ``local GUTs''
\cite{Buchmuller:2004hv,Buchmuller:2005jr,Buchmuller:2005sh,Buchmuller:2006ik}
which inherits certain features of  standard grand unification
\cite{Pati:1974yy,Georgi:1974sy,Fritzsch:1974nn}. Local GUTs  are
specific to certain points in the compact space, while the 4D gauge
symmetry is that of the SM. If matter fields are localized at such
points, they form a complete GUT representation. This applies, in
particular, to a $\boldsymbol{16}$--plet of a local SO(10), which
comprises one generation of the SM matter plus a right--handed
neutrino \cite{Georgi:1975qb,Fritzsch:1974nn},
\begin{equation}
\boldsymbol{16}~=~
(\boldsymbol{3}, \boldsymbol{2})_{1/6} +
(\boldsymbol{\overline{3}}, \boldsymbol{1})_{-2/3}+
(\boldsymbol{\overline{3}}, \boldsymbol{1})_{1/3}+
(\boldsymbol{1}, \boldsymbol{2})_{-1/2} +
(\boldsymbol{1}, \boldsymbol{1})_{1}+
(\boldsymbol{1}, \boldsymbol{1})_{0}
 \;,
\end{equation}
where representations with respect to $\SU3_\mathrm{C}\times\SU2_\mathrm{L}$ are
shown in parentheses and the subscript denotes hypercharge.  On the other hand,
bulk fields are partially projected out and form incomplete GUT multiplets. This
offers an intuitive  explanation for the observed multiplet structure of the SM
\cite{Buchmuller:2004hv,Buchmuller:2005jr,Buchmuller:2005sh,Buchmuller:2006ik}.
This framework is consistent with MSSM gauge coupling unification as long as the
SM gauge group is embedded in a simple local GUT $G_\mathrm{local}\supseteq
\SU5$, which leads to the standard hypercharge normalization.


We find  that the above search strategy, as opposed to  a random scan,
is successful and a considerable fraction of the models with SO(10) and  \E6
local GUT structures pass our criteria. Out of about $ 3\times10^4$ inequivalent
models which involve 2 Wilson lines,  $\mathcal{O}(100)$ are phenomenologically
attractive and can serve as an ultraviolet completion of the MSSM.

\section{MSSM search strategy: local GUTs}

It is well known that with a suitable choice of Wilson lines it is not difficult
to obtain the SM gauge group up to \U1 factors. The real challenge is to get the
correct matter spectrum and  the GUT hypercharge normalization.  To this end, we
base our strategy on the concept  of local GUTs. An  orbifold model is defined
by the orbifold twist, the torus lattice and
the  gauge embedding of the orbifold action,
i.e.\ the gauge shift $V$ and the Wilson lines $W_n$.
We consider only the gauge shifts $V$ which  allow for a local SO(10) or \E6
structure. That is, $V$ are such that the  left--moving momenta $p$
(we use the standard notation, for details see
e.g.~\cite{Forste:2004ie,Kobayashi:2004ya,Buchmuller:2006ik})
satisfying
\begin{equation}\label{eq:LocalGaugeSymmetry}
 p \cdot V ~=~ 0 \mod 1\;,\quad p^2~=~2
\end{equation}
are roots of SO(10) or \E6 (up to extra group factors). Furthermore, the
massless states of the first twisted sector $T_1$  are required to contain  
$\boldsymbol{16}$--plets of SO(10) at the fixed points with SO(10) symmetry
or $\boldsymbol{27}$--plets of \E6 at the fixed points with \E6 symmetry.

Since these massless states from $T_1$ are automatically invariant under the
orbifold action, they all survive in 4D and appear as complete GUT  multiplets.
In the case of \SO{10}, that gives one complete SM generation, while in the case
of \E6 we have $\boldsymbol{27}=\boldsymbol{16}+\boldsymbol{10}+\boldsymbol{1}$
under \SO{10}.  It is thus necessary to decouple all (or part) of the
$\boldsymbol{10}$--plets from the low energy theory.

The Wilson lines are chosen such that the standard model gauge group
is embedded into the local GUT as
\begin{equation}
 G_\mathrm{SM}~ \subset~ \SU5 \subset \SO{10} ~\text{or}~ \E6 \;,
\end{equation}
such that the hypercharge is that of standard GUTs and thus consistent with
gauge coupling unification. The spectrum has certain features of  traditional 4D
GUTs, e.g.\ localized matter fields form complete GUT representations, yet there are
important differences. In particular, interactions generally break GUT relations
since different local GUTs are supported at different fixed points. Also, gauge
coupling unification is due to the fact that the 10D (not 4D) theory is
described by a single coupling.

Our model search is carried out in the \Z6-II orbifold
compactification of the heterotic $\E8\times\E8$ string, which is
described in detail in \cite{Kobayashi:2004ya,Buchmuller:2006ik}. In
this construction, there are 2 gauge shifts leading to a local
\SO{10} GUT \cite{Katsuki:1989cs}, 
\begin{eqnarray}
V^{ \SO{10},1}= &
\left(\tfrac{1}{3},\,\tfrac{1}{2},\,\tfrac{1}{2},\,0,\,0,\,0,\,0,\,0\right)&\left(\tfrac{1}{3},\,0,\,0,\,0,\,0,\,0,\,0,\,0\right) \;,
\nonumber \\
V^{ \SO{10},2 }= &
\left(\tfrac{1}{3},\,\tfrac{1}{3},\,\tfrac{1}{3},\,0,\,0,\,0,\,0,\,0\right)&\left(\tfrac{1}{6},\,\tfrac{1}{6},\,0,\,0,\,0,\,0,\,0,\,0\right) \;,
\label{eq:so10shifts}
\end{eqnarray}
and 2 shifts leading to a local \E6 GUT,
\begin{eqnarray}
 V^{\E6 , 1}= &
\left(\tfrac{1}{2},\,\tfrac{1}{3},\,\tfrac{1}{6},\,0,\,0,\,0,\,0,\,0\right)&\left(0,\,0,\,0,\,0,\,0,\,0,\,0,\,0\right)\;,
\nonumber \\
 V^{ \E6 ,2}= &
\left(\tfrac{2}{3},\,\tfrac{1}{3},\,\tfrac{1}{3},\,0,\,0,\,0,\,0,\,0\right)&\left(\tfrac{1}{6},\,\tfrac{1}{6},\,0,\,0,\,0,\,0,\,0,\,0\right).\label{eq:e6shifts}
\end{eqnarray}
We will focus on these shifts and scan over possible Wilson lines
to get the SM gauge group. The \Z6-II orbifold allows for up to two Wilson lines
of order 2 and for one Wilson line of order 3 (cf.\
[\citen{Kobayashi:1990mi},\citen{Kobayashi:2004ya,Buchmuller:2006ik}]).

The next question is how to get 3 matter generations. The simplest
possibility is to use 3 equivalent fixed points with
$\boldsymbol{16}$--plets \cite{Buchmuller:2004hv} which appear in
models with 2 Wilson lines of order 2. If the extra states are
vectorlike and can be given large masses, the low energy spectrum
will contain 3 matter families. However, this strategy fails since
all such models contain chiral exotic states
\cite{Buchmuller:2006ik}. In the case of \E6, it does not work
either since one cannot obtain
$G_\mathrm{SM}~\subset~\SU5~\subset~\SO{10}~\subset~\E6$
 with 2 Wilson lines of order 2.

The next--to--simplest possibility is to use 2 equivalent fixed
points  which give rise to 2 matter generations. The third
generation would then have to come from other twisted or untwisted
sectors. The appearance of the third family can be linked to the SM
anomaly cancellation. Indeed, the untwisted sector contains part of
a $\boldsymbol{16}$--plet. Then  the simplest options consistent
with the SM anomaly cancellation are that  the remaining matter
either completes the $\boldsymbol{16}$--plet or provides
vector--like partners of the untwisted sector. In more complicated
cases, additional $\boldsymbol{16}$-- or ${\overline
{\boldsymbol{16}}}$--plets can appear. The localized
$\boldsymbol{16}$-- and $\boldsymbol{27}$--plets are true GUT
multiplets, whereas the third or ``bulk''   generation  only has the
SM quantum numbers of an additional $\boldsymbol{16}$--plet. We find
that the above  strategy is successful and  one often gets net 3
families. The other massless states are often vector--like with
respect to the SM gauge group and   can be given large masses
consistent with string selection rules.

In our MSSM search, we focus on models of this type
(although we include all models with 2 Wilson lines in the statistics). These are realized
when 1 Wilson line of order 3 and 1 Wilson line of order 2 are present.
We require that the spectrum contain 3 matter families plus
vector--like states. Furthermore, we discard models in which the SU(5)
hypercharge is anomalous. Although a non--anomalous hypercharge could be
defined, typically it would not have the GUT normalization
and thus would not be consistent with gauge unification.

\section{Results}

Let us now present our results  for models with  the \SO{10} local structure.
For each of the \SO{10}  shifts of Eq.~(\ref{eq:so10shifts}), we follow the steps:
\begin{dingautolist}{"0C0}
 \item Generate Wilson lines $W_3$ and $W_2$.
 \item Identify ``inequivalent'' models.
 \item Select models with $G_\mathrm{SM} \subset \SU5 \subset \SO{10}$.
 \item Select models with three net
 $(\boldsymbol{3},\boldsymbol{2})$.
 \item Select models with non--anomalous $\U1_{Y} \subset \SU5$.
 \item Select models with net 3 SM families + Higgses + vector--like.
\end{dingautolist}
Our results are presented in table \ref{tab:Summary}. The models with the chiral
MSSM matter content are listed in \cite{WebTables:2006ml}.

\begin{table*}[h!]
\centerline{
\begin{tabular}{l||l|l||l|l}
 criterion & $V^{\SO{10},1}$ & $V^{\SO{10},2}$ & $V^{\E6,1}$ & $V^{\E6,2}$\\
\hline
&&&&\\
 \ding{"0C1}  inequivalent models with 2 Wilson lines
  &$22,000$ & $7,800$  &$680$  &$1,700$ \\[0.2cm]
  \ding{"0C2} SM gauge group $\subset$ SU(5) $\subset$ SO(10)
  ({\rm or}~\E6)
  &3563 &1163 &27 &63\\[0.2cm]
  \ding{"0C3} 3 net $(\boldsymbol{3},\boldsymbol{2})$ 
  &1170 &492 &3 &32\\[0.2cm]
  \ding{"0C4} non--anomalous $\U1_{Y}\subset \SU5 $
  &528 &234 &3 &22\\[0.2cm]
  \ding{"0C5}  spectrum $=$ 3 generations $+$ vector-like
  &128 &90 &3 &2
  \\
\hline
\end{tabular}
}
\caption{Statistics of \Z6-II orbifolds based on the shifts
$V^{\SO{10},1},V^{\SO{10},2},V^{\E6,1},V^{\E6,2}$ with two Wilson lines.
\label{tab:Summary} }
\end{table*}

 Before continuing further, we make a few comments.   In order to
obtain the models listed under points  \ding{"0C2}- \ding{"0C5}, we
generate all possible Wilson lines along the lines of
Refs.~\cite{Giedt:2000bi} and \cite{Nilles:2006np}. However, due to
the rapid growth in computing time, generating all inequivalent
models is not possible using these tools.  Thus the inequivalent
models under point \ding{"0C1} have been generated by exploiting
symmetries of the gauge lattice along the lines discussed in
\cite{Wingerter:thesis}.
Two models are considered ``equivalent'' if they have identical spectra
with respect to non--Abelian gauge groups and have the same number
of non--Abelian singlets. Thus, models differing only in U(1)
charges are treated as equivalent. Further ambiguities arise in
certain cases when  $\U1_Y$ can be defined in different ways.
In addition, some models differ only by the
localization of states on the different fixed points. We know that
these ambiguities occur and it is possible that in some cases Yukawa
couplings are affected. Hence our criterion may underestimate the
number of truly inequivalent models.

In the \E6 case, we consider the SM embedding
\begin{equation}
G_\mathrm{SM}~\subset~\SU5~\subset~\SO{10}~\subset~\E6 \;.
\end{equation}
Again, models with 2 Wilson lines of order 2 fail and  the  analysis
proceeds similarly to the \SO{10} case. These results are also
presented in table \ref{tab:Summary}.\footnote{In the analysis of 
\cite{Dienes:2006ut} looking at non-supersymmetric heterotic 
string vacua, about 10\% of the models scanned contained the SM 
gauge group. Our result (step 3) is comparable.}

It is instructive to compare our model scan to others. In certain types of intersecting
D--brane models, it was found that the probability of obtaining  the SM gauge
group and three generations of quarks and leptons, while allowing for chiral
exotics, is $10^{-9}$ \cite{Gmeiner:2005vz,Douglas:2006xy}.
The criterion which comes closest to the requirements imposed in
\cite{Gmeiner:2005vz,Douglas:2006xy} is \ding{"0C3}.
We find that within our sample the corresponding probability
is 5\,\%.

In \cite{Dijkstra:2004cc,Anastasopoulos:2006da}, orientifolds of
Gepner models were scanned for chiral MSSM matter spectra, and it
was found that the fraction of such models is $10^{-14}$. In our set
of models, the corresponding probability, i.e.\ the fraction of
models passing criterion \ding{"0C5}, is of order 1\,\%. Note also
that, in all of our models, hypercharge is normalized as in standard
GUTs and thus  consistent with  gauge coupling unification.

This comparison shows that our sample of heterotic orbifolds is unusually
``fertile'' compared to other constructions.
 The probability of
finding something close to the MSSM is much higher than that in other patches of the
landscape analyzed so far.
It would be interesting to extend these results to other regions of the
landscape where  promising models exist
\cite{Braun:2005nv,Bouchard:2005ag,Kim:2006hv,Cleaver:1998sa} (see also 
\cite{Blumenhagen:2006ux}).

\section{Towards realistic string models}

The next step on the path towards realistic models is the decoupling of the
vector--like extra matter  $\{ x_i \}$. The mass terms for such states are
provided by the superpotential
\begin{equation}\label{eq:decoupling}
 W~=~ x_i\, \bar x_j\, \langle s_a\, s_b\dots \rangle \;,
\end{equation}
where $s_a,s_b,\dots$ are SM singlets.
Some singlets are required to get large (close to $M_\mathrm{str}$) VEVs in
order to cancel the Fayet--Iliopoulos (FI) term of an anomalous U(1). The
supersymmetric field configurations are quite complicated and generally there
are vacua in which all or most of the SM singlets get large  VEVs.  This breaks
many of the gauge group factors, such that the low energy gauge group can be
$G_\mathrm{SM}$ up to a hidden sector,
\begin{equation}
G_\mathrm{SM}
 \times G_\mathrm{hidden}\;,
\end{equation}
where the SM matter is neutral under $G_\mathrm{hidden}$.
Furthermore, if the relevant Yukawa couplings are allowed by string
selection rules, this  makes the vector--like matter heavy; thus it
decouples from the low energy theory.
We note that there are in general several pairs of Higgs doublets with a matrix
of $\mu$--like mass terms, for which we  require only one small eigenvalue.
\footnote{To get a pair of massless Higgs doublets usually requires fine-tuning
in the VEVs of the SM singlets such that the mass matrix for the $({\bf 1}, {\bf
2})_{-1/2}, ({\bf 1}, {\bf 2})_{1/2} $ states gets a zero eigenvalue. This is
the notorious supersymmetric $\mu$--problem. The fine-tuning can be ameliorated if the vacuum
respects certain (approximate) symmetries \cite{Kim:1983dt,Giudice:1988yz}.
}

Clearly, one cannot switch on the singlet VEVs at will.  Instead one
has to ensure that they are consistent with supersymmetry.
Supersymmetry requires vanishing of the $F$-- and $D$--terms. The
number of the  $F$--term equations equals the  number of complex
fields $s_a$, therefore there are in general non--trivial singlet
configurations with vanishing $F$--potential. The $D$--terms can be
made zero by complexified gauge transformations \cite{Ovrut:1981wa}
if each field enters a gauge invariant monomial
\cite{Buccella:1982nx}. Thus, to ensure that the decoupling of
exotics is consistent with supersymmetry, one has to show that all
SM singlets appearing in the mass matrices for the exotics enter
gauge  invariant monomials involving only SM singlets and carrying
anomalous charge. In this letter, we assume that the relevant
singlets develop large supersymmetric VEVs.

In the process of decoupling, the vector--like states can mix with
the localized $\boldsymbol{16}$-- and $\boldsymbol{27}$--plets (if
it is allowed by the SM quantum numbers) such that the physical
states at low energies are \emph{neither} localized \emph{nor}
``true'' GUT multiplets. Nevertheless, it is clear that whatever the
mixing, in the end exactly 3 SM families will be left, if the mass
matrices have maximal rank.

To show that the decoupling of exotics is
consistent with string selection rules is a technically involved and
time consuming issue. In order to simplify the task and to reduce
the number of models, we first impose an additional condition.  We
require that the models possess a renormalizable top--Yukawa coupling
as motivated by phenomenology.
Then  we consider only superpotential couplings up to order 8.  
Thus, the next two steps in our
selection procedure are:
\begin{dingautolist}{"0C6}
  \item Select models with a heavy top.
  \item Select models in which the exotics decouple at order 8.
\end{dingautolist}
First, we
require a renormalizable $\mathcal{O}(1)$ Yukawa coupling
  $(\boldsymbol{3}, \boldsymbol{2})_{1/6}\,  (\boldsymbol{\bar 3},
\boldsymbol{1})_{-2/3} \, (\boldsymbol{1}, \boldsymbol{2})_{1/2}$, i.e. one of the
following types
\begin{equation}
 U\,U\,U\;,\quad U\,T\,T\;,\quad T\,T\,T \;,
\end{equation}
where $U$ and $T$ denote generic untwisted and twisted fields, respectively.
The $U\,U\,U$ coupling is given by the gauge coupling,  $U\,T\,T$ is a local
coupling and thus is unsuppressed, while the $T\,T\,T$ coupling is significant
only when the twisted fields are localized at the same fixed point.
We discard models in which the above couplings are absent or
suppressed.

In the next step \ding{"0C7}, we select models in which the  mass
matrices for the exotics (cf.\ Eq.~\eqref{eq:decoupling}) have a
maximal rank such that no exotic states appear at low energies.
Here, we consider only superpotential couplings up to order 8 and
for this analysis we assume that all relevant singlets can obtain
supersymmetric vevs.\footnote{We also address to some extent the
question of $D$--flatness. In many of the models, we find that all
SM singlets enter gauge invariant monomials.  A full analysis of
this issue is deferred to a subsequent publication.}
We find that a significant fraction of our models passes
requirements \ding{"0C6} and \ding{"0C7} (see
table~\ref{tab:Summary2} and for further details
\cite{WebTables:2006ml}). In particular, we identify 93 models that
can serve as an ultraviolet completion of the MSSM in string theory.

\begin{table*}[t!]
\centerline{
\begin{tabular}{l||l|l||l|l}
 criterion & $V^{\SO{10},1}$ & $V^{\SO{10},2}$ & $V^{\E6,1}$ & $V^{\E6,2}$\\
\hline
&&&&\\
  \ding{"0C6}  heavy top
   &72 &37 &3 &2\\[0.2cm]
  \ding{"0C7}  exotics decouple at order 8
   &56 & 32 & 3 & 2
  \\
\hline
\end{tabular}
}
\caption{A subset of the MSSM candidates.
\label{tab:Summary2} }
\end{table*}

To verify  whether an MSSM candidate is consistent with phenomenology requires
addressing several questions. The most important issues include
\begin{itemize}
 \item realistic flavour structures,
 \item absence of fast proton decay,
 \item hierarchically small supersymmetry breaking.
\end{itemize}

A model that passes all
of our criteria \ding{"0C2}--\ding{"0C7} and comes very close to the
supersymmetric standard model has been presented in
\cite{Buchmuller:2005jr,Buchmuller:2006ik}.   In our scan, we obtain many
comparable models.   In what follows, we
substantiate this statement by studying a specific example, leaving a complete
survey for future work.

\subsection*{Example}

The model is based on the gauge shift
\begin{eqnarray}
V^{ {\rm SO(10)},1} & =
&\left(\tfrac{1}{3},\,-\tfrac{1}{2},\,-\tfrac{1}{2},\,0,\,0,\,0,\,0,\,0\right)\left(\tfrac{1}{2},\,-\tfrac{1}{6},\,-\tfrac{1}{2},\,-\tfrac{1}{2},\,-\tfrac{1}{2},\,-\tfrac{1}{2},\,-\tfrac{1}{2},\,\tfrac{1}{2}\right)
\;.
\end{eqnarray}
where we have added an \E8$\times$\E8 lattice vector to simplify
computations.  The Wilson lines are chosen as
\begin{eqnarray}
W_{2} & = &\left(\tfrac{1}{4},\,-\tfrac{1}{4},\,-\tfrac{1}{4},-\tfrac{1}{4},\,-\tfrac{1}{4},\,\tfrac{1}{4},\tfrac{1}{4},\,\tfrac{1}{4}\right)\left(1,\,-1,\,-\tfrac{5}{2},\,-\tfrac{3}{2},\,-\tfrac{1}{2},\,-\tfrac{5}{2},\,-\tfrac{3}{2},\,\tfrac{3}{2}\right) \;, \nonumber \\
W_{3} & = &\left(-\tfrac{1}{2},\,-\tfrac{1}{2},\,\tfrac{1}{6},\tfrac{1}{6},\,\tfrac{1}{6},\,\tfrac{1}{6},\tfrac{1}{6},\,\tfrac{1}{6}\right)\left(\tfrac{10}{3},\,0,\,-6,\,-\tfrac{7}{3},\,-\tfrac{4}{3},\,-5,\,-3,\,3\right)\;.
\end{eqnarray}
The standard SU(5)  hypercharge generator is given by
\begin{eqnarray}
\mathsf{t}_{Y}   & = &\left(0,\,0,\,0,\,-\tfrac{1}{2},\,-\tfrac{1}{2},\,\tfrac{1}{3},\,\tfrac{1}{3},\,\tfrac{1}{3}\right)\left(0,\,0,\,0,\,0,\,0,\,0,\,0,\,0\right)
\;.
\end{eqnarray}
The gauge group after compactification is
\begin{equation}
 G_\mathrm{SM} \times \SO8 \times \SU2 \times \U1^7 \;,
\end{equation}
while the massless spectrum is given in table \ref{spectrum}.
\begin{table}[!h]
\begin{center}
\begin{tabular}{|rl|c|rl|c|c|rl|c|}
\cline{1-6}\cline{8-10}
  \#  &  irrep                                                         & label           &  \#  & anti-irrep                                         & label       &&  \#  & irrep & label\\
\cline{1-6}\cline{8-10}
  3 & $(  {\bf 3},  {\bf 2};  {\bf 1},  {\bf 1})_{1/6}$             & $q_i$           &    & $\phantom{I^{I^I}}$                                &             &&  4 & $(  {\bf 1},  {\bf 2};  {\bf 1},  {\bf 1})_{0}$ & $m_i$\\
  8 & $(  {\bf 1},  {\bf 2};  {\bf 1},  {\bf 1})_{-1/2}$            & $\ell_i$
          &  5 & $(  {\bf 1},  {\bf 2};  {\bf 1},  {\bf 1})_{1/2}$  &
          $\bar{\ell}_i$ &&  2 & $(  {\bf 1},  {\bf 2};  {\bf 1},  {\bf
          2})_{0}$ & $m_i'$\\
  3 & $(  {\bf 1},  {\bf 1};  {\bf 1},  {\bf 1})_{1}$               & $\bar{e}_i$     &    &                                                    &             && 47 & $(  {\bf 1},  {\bf 1};  {\bf 1},  {\bf 1})_{0}$ & $s_i$\\
  3 & $(  {\bf \overline{3}},  {\bf 1};  {\bf 1},  {\bf 1})_{-2/3}$ & $\bar{u}_i$     &    &                                                    &             && 26 & $(  {\bf 1},  {\bf 1};  {\bf 1},  {\bf 2})_{0}$ & $h_i$\\
  7 & $(  {\bf \overline{3}},  {\bf 1};  {\bf 1},  {\bf 1})_{1/3}$  & $\bar{d}_i$     &  4 & $(  {\bf 3},  {\bf 1};  {\bf 1},  {\bf 1})_{-1/3}$ & $d_i$       &&  9 & $(  {\bf 1},  {\bf 1};  {\bf 8},  {\bf 1})_{0}$ & $w_i$\\
\cline{1-6}\cline{8-10}
  4 & $(  {\bf 3},  {\bf 1};  {\bf 1},  {\bf 1})_{1/6}$             & $v_i$           &  4 & $(  {\bf \overline{3}},  {\bf 1};  {\bf 1},  {\bf 1})_{-1/6}$ & $\bar{v}_i$     &
  \multicolumn{3}{c}{$\phantom{I^{I^I}}$} 
  \\
 20 & $(  {\bf 1},  {\bf 1};  {\bf 1},  {\bf 1})_{1/2}$             & $s^+_i$         & 20 & $(  {\bf 1},  {\bf 1};  {\bf 1},  {\bf 1})_{-1/2}$            & $s^-_i$         &
\multicolumn{3}{c}{$\phantom{I^{I^I}}$} 
\\
  2 & $(  {\bf 1},  {\bf 1};  {\bf 1},  {\bf 2})_{1/2}$             & $\tilde{s}^+_i$ &  2 & $(  {\bf 1},  {\bf 1};  {\bf 1},  {\bf 2})_{-1/2}$            & $\tilde{s}^-_i$ &
  \multicolumn{3}{c}{$\phantom{I^{I^I}}$} 
  \\
\cline{1-6}
\end{tabular}
\caption{Massless spectrum. The quantum numbers are shown with respect to
$\SU3_\mathrm{C} \times \SU2_\mathrm{L} \times \SO8 \times \SU2$, the hypercharge is
given by the subscript.}
\label{spectrum}
\end{center}
\end{table}

Renormalizable Yukawa couplings involving the SM fields are shown in
table \ref{inter}. The top Yukawa coupling comes from the $U\,U\,U$
interaction $q_i \bar{\ell}_j \bar{u}_k$, which allows us to
identify the right--handed top, the up--type Higgs doublet and the
quark doublet of the third generation. (Here we denote the leptons
and Higgses collectively by $\ell_i, \bar \ell_i$.)  Other
renormalizable interactions  $q_i\, \ell_j\, \bar{d}_k$ and
$\bar{e}_i\, \ell_j\, \ell_k$ can produce the down--type quark  and
lepton masses as well as lepton number violating interactions. What
happens precisely depends on the form of the matrix of $\mu$-like
mass terms for the vector--like states  and, thus, on the vacuum
configuration. We note that, due to the absence of  the $\bar{u}_i
\bar{d}_j \bar{d}_k$ operator, the proton is stable at this level.

\begin{table}[!h]
\begin{center}
\begin{tabular}{|l||l|l|l|l|}
\hline
coupling &
$q_i\, \bar{\ell}_j\, \bar{u}_k$ &
$\bar{u}_i\, \bar{d}_j\, \bar{d}_k$ &
$q_i\, \ell_j\, \bar{d}_k$ &
$\bar{e}_i\, \ell_j\, \ell_k$ \\
\hline
\#
& 1
& 0
& 4
& 4 \\
\hline
\end{tabular}
\caption{Renormalizable interactions involving the SM fields.}
\end{center}
\label{inter}
\end{table}

The model has three generations of SM matter plus vector--like
exotics. Once the SM singlets $s_i$ get VEVs, the gauge group
reduces to
\begin{equation}
G_\mathrm{SM} \times G_\mathrm{hidden} \;,
\end{equation}
where $ G_\mathrm{hidden}= \SO8 \times \SU2$. At the same time, the
vector--like states get large masses. We have checked that the rank
of all the  mass matrices is maximal, such that the exotics do
decouple (assuming all singlets acquire supersymmetric vevs). Below
we present most of them. An entry $s^n$ indicates that the coupling
appears  first when $n$ singlets are involved. Each entry usually contains
many terms and involves different singlets as well as coupling
strengths.
\begin{eqnarray}
\lefteqn{
\begin{array}{ll}\!\!\!\!\!\!\!\!\!
\begin{array}{l}
 \mathcal{M}_{d \bar d}~=~\left(
\begin{array}{ccccccc}
s^6 & s^6 & s^3 & s^6 & s^6 & s^1 & s^1\\
s^6 & s^6 & s^3 & s^6 & s^6 & s^1 & s^1\\
s^3 & 0 & 0 & s^3 & 0 & s^6 & s^6\\
s^6 & s^3 & 0 & s^6 & s^3 & s^6 & s^6\\
\end{array}
\right) \;,\\ \\
\mathcal{M}_{m' m'}=
\left(
\begin{array}{cc}
s^1   & s^5 \\
s^5 & s^1   \\
\end{array}
\right)\;,
\end{array}
&
\mathcal{M}_{\ell \bar\ell}~=~\left(
\begin{array}{ccccc}
s^3 & s^1 & s^1 & s^1 & s^1\\
s^1 & s^3 & s^3 & s^3 & s^3\\
s^1 & s^3 & s^3 & s^3 & s^3\\
s^1 & s^3 & s^3 & s^3 & s^3\\
s^1 & s^3 & s^3 & s^3 & s^3\\
s^1 & s^3 & s^6 & s^6 & s^3\\
s^4 & s^2 & s^6 & s^2 & s^2\\
s^4 & s^2 & s^6 & s^2 & s^2\\
\end{array}
\right) \;, \nonumber
\end{array}
}\nonumber\\
\lefteqn{
\mathcal{M}_{v \bar v}~=~ \left(
\begin{array}{cccc}
s^5 & s^5 & s^5 & s^5\\
s^5 & s^5 & s^5 & s^5\\
s^6 & s^6 & s^1 & s^5\\
s^6 & s^6 & s^5 & s^1\\
\end{array}
\right) \;,\quad
\mathcal{M}_{m  m}= \left(
\begin{array}{cccc}
0   & s^5 & s^6 & s^6\\
s^5 & 0   & s^6 & s^6\\
s^6 & s^6 & 0 & s^5\\
s^6 & s^6 & s^5 & 0\\
\end{array}
\right) \;.}
\nonumber
\end{eqnarray}
Similarly, the mass matrices for $s_i^\pm$ and  $\tilde s_i^\pm$
have a maximal rank. The $d\bar d$ mass matrix is 4$\times$7 such
that there are 3 massless $\bar d$ states. The $\ell \bar\ell$ mass
matrix is 8$\times$5, so there are 3 lepton doublets. By choosing a
special vacuum configuration one can reduce the rank of the $\ell
\bar \ell$ mass matrix to 4 such that there is a pair of massless
Higgs doublets. (This is just the supersymmetric
``$\mu$--problem'').  Thus we end up with the exact MSSM spectrum.

We have checked that the required vacuum configuration is $D$--flat. That is,
one can assign large VEVs to the singlets without inducing the $D$--terms.
Since the number of the $F$--term equations equals the number of the field  variables,
there are generally non--trivial solutions to $F=0$. 
Then, using complexified gauge transformations, one can make the
$F$-- and $D$--terms vanish simultaneously. Such supersymmetric vacua
would  correspond to isolated solutions in field space. 
Although we expect such solutions to exist,  their explicit
form remains undetermined and will be studied elsewhere.

Finally, the  model  allows us  to define a  suitable
$B-L$ generator which leads to the standard charges
for the SM matter,
\begin{equation}
\mathsf{t}_{B-L}
~=~\left(1,\,1,\,0,\,0,\,0,\,-\tfrac{2}{3},\,-\tfrac{2}{3},\,-\tfrac{2}{3}\right)\left(2x
- \tfrac{1}{2},\,\tfrac{1}{2},\,0,\,x,\,x,\,0,\,0,\,0\right) \;,
\end{equation}
with arbitrary $x$. An interesting feature is that the spectrum
contains a pair of fields which have $B-L$ charges $\pm 2$. If $B-L$
gauge symmetry is broken by VEVs of these fields, the matter parity
(or family reflection symmetry \cite{Dimopoulos:1981zb,Dimopoulos:1981dw})
$(-1)^{3(B-L)}$ is conserved and proton decay is suppressed.

\section{Conclusion}

We have analyzed the heterotic $\E8\times\E8$ string compactified on
a \Z6-II orbifold, allowing for up to two discrete Wilson lines.
Employing a search strategy based on the concept of local GUTs, we
have obtained about $3\times 10^4$ inequivalent models. Almost 1\,\%
of these models have the gauge group and the chiral matter content
of the MSSM. This result shows that orbifold compactifications of
the  heterotic string  considered here correspond to a particularly fertile region in the
landscape  and  the probability of
getting something close to the MSSM is  significantly higher than
that in other constructions.

The most important outcome of our scan is the construction of
$\mathcal{O}(100)$ models consistent with the MSSM gauge group and
matter content, amended by a hidden sector. A detailed
phenomenological analysis of these models is in progress.

\subsubsection*{Acknowledgments}

We would like to thank A.~N.~Schellekens and W.~Taylor for
correspondence. This work was partially supported by the European
Union 6th Framework Program MRTN-CT-2004-503369 ``Quest for
Unification'' and MRTN-CT-2004-005104 ``ForcesUniverse''. S.~Raby
and A. Wingerter received partial support from DOE grant
DOE/ER/01545-870, and would like to thank the KITP for their
hospitality during the completion of this work. This research was
supported in part by the National Science Foundation under Grant No.
PHY99-07949.

\end{document}